\documentclass[twocolumn,showpacs]{revtex4}
\usepackage{graphics,epsfig,graphicx}
\usepackage{color}
\usepackage{makeidx}
\usepackage{latexsym}

\begin{document}


\title{Photon correlation vs.\ interference of single-atom fluorescence in a half-cavity}

\author{Fran\c{c}ois Dubin$^1$, Daniel Rotter$^1$, Manas Mukherjee$^1$, Carlos Russo$^1$,
J\"{u}rgen Eschner$^2$, and Rainer Blatt$^{1\dag}$}

\affiliation{$^1$Institute for Experimental Physics, University of Innsbruck,
Technikerstr.\ 25, A-6020 Innsbruck, Austria \\
$^2$ICFO -- Institut de Ci\`{e}ncies Fot\`{o}niques, 08860
Castelldefels (Barcelona), Spain.\\
$^{\dag}$Institute for Quantum Optics and Quantum Information of
the Austrian Academy of Sciences, 6020 Innsbruck, Austria}

\date{\today }


\begin{abstract}
Photon correlations are investigated for a single laser-excited
ion trapped in front of a mirror. Varying the relative distance
between the ion and the mirror, photon correlation statistics can
be tuned smoothly from an antibunching minimum to a bunching-like
maximum. Our analysis concerns the non-Markovian regime of the
ion-mirror interaction and reveals the field establishment in a
half-cavity interferometer.
\end{abstract}

\maketitle


Experiments with laser-cooled trapped ions have provided important
contributions to the understanding of quantum phenomena. A single trapped ion
is in fact a model system whose internal and external degrees of freedom can be
controlled at the quantum level: non-classical motional states such as Fock
states and quadrature-squeezed states have been successfully engineered with a
single Be$^+$ ion \cite{squeezed-state}; the internal levels of trapped ions
have been coherently manipulated by sequences of laser pulses, and have been
entangled with the motional state, leading to the preparation of
Schr\"{o}dinger cat states \cite{Cat-state} and to multi-ion entangled states
for quantum information processing \cite{Q-bite}.

The internal dynamics of a laser-driven single ion or atom is well
characterized by the statistical analysis of the measured stream
of fluorescence photons, namely by the second order correlation
function $G^{(2)}(T)$ \cite{MW}, i.e.\ the frequency of time
intervals of length $T$ between detected photons. For a single
atom trapped in free space, this correlation function exhibits
sub-Poissonian statistics and violates the Cauchy-Schwarz
inequality, i.e.\ $G^{(2)}(0) < G^{(2)}(T)$. More precisely,
$G^{(2)}(T)$ exhibits a minimum at $T=0$ which indicates the
quantum nature of photon emission, or the projective character of
photon detection. This is defined as anti-bunching \cite{KDM,
Walther}. On the contrary, for a large ensemble of atoms the
emitted radiation exhibits classical bunching \cite{HBT}
fulfilling $G^{(2)}$(0)$\geq$ $G^{(2)}(T)$. A smooth transition
from anti-bunching to bunching has recently been observed in a
high-Q resonator when increasing the number of interacting atoms
\cite{Hennrich}.

The second order correlation function can be viewed as
representing the (average) dynamics of the observed system
conditioned on the emission of a photon at time $T=0$. While
$G^{(2)}$ thereby draws on the photon character of the emitted
light, it is the wave character which is responsible for
interference phenomena, in particular for QED effects in
resonators. In this letter, we examine the interplay of photon
detection and wave interference in a simple cavity QED experiment,
by measuring the second order photon correlation for a single
trapped Ba$^+$ ion in a half-cavity interferometer. In this set-up
part of the resonance fluorescence of the laser-excited ion is
retro-reflected by a mirror at a distance $L$ and focussed back
onto its source. Earlier experiments with our system revealed
back-action of the interferometer on the emitting atom such as
modification of its decay rate \cite{Eschner} and energy shifts of
the excited state \cite{Wilson}; even mechanical action was
observed \cite{Bushev}. Such effects intrinsically pertain to the
interference caused by the mirror.
On the other hand, the mirror induces a time delay $\tau=2L/c$,
needed for photons to return to the ion's position. When $\tau$ is
negligible on the time scale of the atomic dynamics, the modified
decay rate and energy shift correspond to the "low-Q" regime of
cavity QED \cite{Milonni}. Here we investigate a different regime,
when $\tau$ is comparable to the spontaneous emission lifetime.
This characterizes a non-Markovian situation, where retardation
and memory effects play a major role: the emitted photon projects
the atom, and interference can only be established after the delay
time $\tau$, when the atomic dynamics have already evolved
significantly \cite{Motion}. This problem was first discussed
theoretically by Cook and Milonni \cite{MC}, then by Alber
\cite{Alber}, and recently by Dorner and Zoller \cite{DZ} with a
particular emphasis on our experimental conditions. Our study is,
to our knowledge, the first single-atom implementation of such a
system.

We report measurements for two ion-mirror distances, $L=67$~cm and
90~cm, and find them in quantitative agreement with theoretical
predictions. Depending on the exact position of the mirror, which
we vary on the nanometer scale, the $G^{(2)}$ function shows
radically different behaviour. In particular, we observe how the
interference in the mode reflected by the mirror sets in with the
retardation time $\tau$. At a more general level, this corresponds
to a sudden transition in the dynamics of the atom-cavity system
from a regime where which-way information is present to the regime
where interference is established. Moreover, through varying $L$,
the value of $G^{(2)}(0)$ for our single atom can be tuned from an
anti-bunching minimum to a bunching-like maximum.

\begin{figure}
\begin{center}
\includegraphics[width=0.45\textwidth]{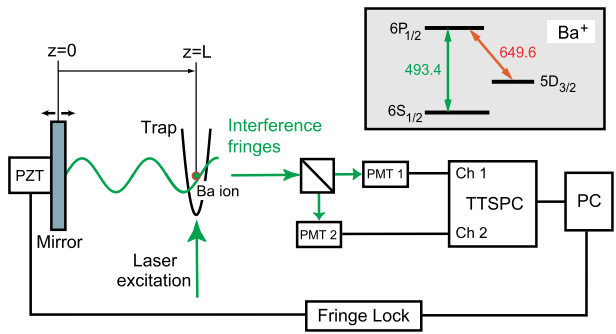}
\caption{A single $^{138}$Ba$^{+}$ ion in a Paul trap (parabola)
is continuously laser-excited. A lens (not shown) and a mirror at
distance $L$, mounted on piezo-actuators (PZT), focus back part of
the fluorescence onto the ion. Green (493~nm) photons are detected
by two photomultipliers (PMT 1 and 2) and their arrival times are
correlated with 100 ps temporal resolution (TTSPC: Time Tagged
Single Photon Counting). A slow electronic servo loop (fringe
lock) stabilises the average photocurrent and thereby permits
control of the distance $L$ between the ion and the mirror with
better than 10~nm precision.}\label{fig1}
\end{center}
\end{figure}

The schematic experimental set-up and the relevant partial level
scheme of $^{138}$Ba$^+$ are shown in Fig.1. The ion is
continuously driven and cooled by two narrow-band tunable lasers
at 493~nm (green) and 650~nm (red) exciting the
S$_{1/2}$--P$_{1/2}$ and P$_{1/2}$--D$_{3/2}$ transitions,
respectively. Laser frequencies are close to resonance and
intensities are set below saturation. A fraction $\epsilon$ of the
green fluorescence photons is reflected by a distant mirror and
focussed back onto the ion. We analyse the $G^{(2)}$ correlation
function of the 493~nm light in the observation channel opposite
to the mirror. This light has two components, the direct and the
reflected part of the radiation scattered by the ion, with a time
delay $\tau$ between them. For very low laser intensities, when
all scattering is elastic, the resulting interference of these
components is observed with up to 72$\%$ visibility \cite{Eschner}
into that mode. In the measurements presented here we use slightly
higher laser excitation rate, whereby the contrast reduces to
around 50$\%$.
The interference signal can be viewed as a consequence of the
standing wave which forms in the mirror mode and which leads to
inhibited and enhanced detection of resonance fluorescence photons
\cite{Eschner}. The signal varies with the ion-mirror distance $L$
as sin$^2(k_{fl}L)$, where $k_{fl}$ is the momentum of photons
emitted at 493~nm. A fringe minimum corresponds to the ion being
located at a node of the standing wave, i.e.\ $k_{fl}L=n\pi$ ($n$
being an integer); the maximum corresponds to
$k_{fl}L=(n+\frac{1}{2})\pi$, i.e.\ to the ion being at an
antinode.

We note that on average there are less than 10$^{-3}$ photons in
the mode volume between the ion and the mirror. This gives rise to
one of the remarkable features of this experiment, that the
interference is created by partial waves corresponding to the same
individual photon, while at the same time the detection of these
photons reveals dynamical information and state projection of the
atom.

We now study the second order correlation for arrival times of green photons.
First we recall the main theoretical results of Ref.\cite{DZ}, restricting the
treatment to the S$_{1/2}$ and P$_{1/2}$ levels. As shown in Fig.\ref{fig1}, we
label the mirror-ion-detector axis as $z$, set the mirror position at $z=0$ and
the trap center at $z=L$. Neglecting the motion of the ion in the trap, the
field operator for green photons in the mirror mode reads at $z=L$
\begin{eqnarray}
E_m(L,t) = \frac{\epsilon\Gamma}{2} \frac{i\hbar}{d}
e^{-i\omega_L t} [\sigma^-(t) \theta(t) \nonumber\\
-e^{i\omega_L \tau} \sigma^-(t-\tau)\theta(t-\tau)] +N_v(t)~,
\label{eq1}
\end{eqnarray}
where $\theta(t)$ is a step function centered at $t=0$, $\Gamma$ is the
free-space decay rate of the P$_{1/2}$ to S$_{1/2}$ transition, and $d$ its
dipole oscillator strength. $\sigma^-$ denotes the lowering operator from
$|{\rm P}_{1/2}\rangle$ to $|{\rm S}_{1/2}\rangle$ and $\omega_L$ the laser
frequency. $N_v$ is the source free part of the mirror field, i.e. the input
state in the language of input-output theory \cite{GZ}. In Eq.(\ref{eq1}) the
interaction picture with respect to the free part of the Hamiltonian is used,
operators become time dependent, and we turn into a frame rotating at the laser
frequency, e.g.\ $\sigma^{-}(t) \rightarrow \sigma^-(t) e^{-i\omega_Lt}$.
Including proper commutation rules between input and output states of the
field, the second order time correlation function in the mirror mode,
$G_m^{(2)}(t,t+T)=\langle E^{\dag}_m(L,t)E^{\dag}_m(L,t+T) E_m(L,t+T)E_m(L,t)
\rangle$, reads
\begin{eqnarray}
G_m^{(2)}(t,t+T)\propto\|\sigma^-(t+T)\sigma^-(t)\nonumber\\
+e^{2i\omega_L\tau}\sigma^-(t+T-\tau)\sigma^-(t-\tau)\nonumber\\
-{\cal T}_{\hookleftarrow} e^{i\omega_L\tau} \sigma^-(t+T-\tau)\sigma^-(t)\nonumber\\
-e^{i\omega_L\tau}\sigma^-(t+T)\sigma^-(t-\tau)|i\rangle\|^2~, \label{eq3}
\end{eqnarray}
where $|i\rangle$ denotes the initial state of the system, i.e.\
the ion in the ground state $|{\rm S}_{1/2}\rangle$ and the mirror
mode in the vacuum state. The different contributions in
Eq.(\ref{eq3}) are interpreted as follows: the first term
corresponds to the detection of two photons directly emitted
towards the detectors and separated by a time interval $T$; in the
second term, these photons are both reflected by the mirror
(therefore delayed by $\tau$) before detection. The two last
contributions describe possible detection of either first a
directly emitted photon and then a second one after its reflection
on the mirror (third term), or vice-versa (fourth term). In the
former case, for $T<\tau$ causality is ensured by ${\cal
T}_{\hookleftarrow}$ which enforces the time ordering of the two
operators on its right hand side. These must be arranged
chronologically from right to left and have to be commuted if they
are not. Consequently, in Eq.(\ref{eq3}) different contributions
interfere. The first two terms induce anti-bunching around $T=0$
while the two others may counteract this usual behavior. As we
show below, the weight of each component strongly depends on the
actual position of the ion, i.e.\ wether it is located at a node
or at an anti-node of the mirror mode. Finally, from
Eq.(\ref{eq3}) one obtains in the steady-state limit
($t\rightarrow\infty$)
\begin{eqnarray}
G_m^{(2)}(T) \propto\ |2b_{{\rm P}_{1/2}}(T)\cos(2k_{fl}L)\nonumber\\
-b_{{\rm P}_{1/2}}(|T-\tau|)-b_{{\rm P}_{1/2}}(T+\tau)|^2 \label{eq4}
\end{eqnarray}
where $b_{{\rm P}_{1/2}}$ denotes the occupation amplitude of the P$_{1/2}$
level. In principle, it should be evaluated including the mirror induced
modifications of decay rate and energy value of the P$_{1/2}$ state
\cite{Eschner, Wilson, Bushev}. Nevertheless, the mirror back-action can be
neglected for the current analysis, with $\epsilon$ being on the order of
1.5$\%$. Then $b_{{\rm P}_{1/2}}$ is deduced from the density matrix time
evolution considering a single Ba$^{+}$ ion trapped in free space. Note that
all 8 electronic sub-levels need to be accounted for in order to accurately
reproduce the exact shape of the measured correlations \cite{Toschek}.

\begin{figure}
\includegraphics[width=6cm]{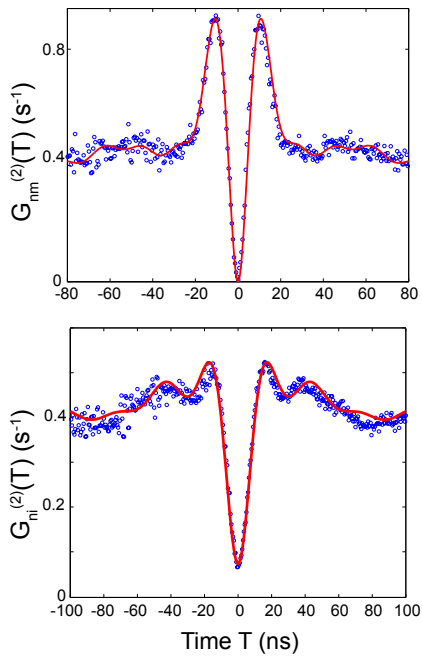}
\caption{{\it Top}: Measured second order correlation function
without mirror, $G^{(2)}_{nm}$ (circles) and its simulation
calculated from 8-level Bloch equations (line). {\it Bottom}:
Correlation function for non-interfering ion and mirror image,
$G^{(2)}_{ni}$, for $\tau=4.5$~ns. The line is the sum of three
correlation functions as explained in the text. For the measured
curves we evaluate the time intervals between all pairs of
detected photons using a 500~ps time bin width, and then divide
the data by the total integration time (several hours) after
background subtraction.} \label{fig2}
\end{figure}

In the top panel of Fig.\ref{fig2}, we present the correlation
function in absence of the mirror, $G^{(2)}_{nm}$. It is obtained
using the set-up depicted in Fig.\ref{fig1}, but with the mirror
blocked. The measurement exhibits the characteristic anti-bunching
at short time, with a null rate of coincidences, $G^{(2)}_{nm}(0)
\simeq 0$. It is accurately reproduced by our simulations which do
not require any fitting parameter, only experimental conditions
such as laser powers and detunings \cite{Toschek}. The lower panel
of Fig.\ref{fig2} shows the correlation function when the mirror
is included, but without overlapping the reflected field with its
source; ion and mirror image are then spatially distinct, and
there is no interference. The signal, $G^{(2)}_{ni}(T)$,
corresponds to three synchronous but non-interfering sources,
shifted in time by $\pm\tau$. The expected contributions to this
signal are the moduli squares of the three terms in
Eq.(\ref{eq4}), without the cosine dependance, i.e. without
interference. As shown by the full line, their sum accurately
reproduces our measurements. In the following this signal is used
as a reference: in the model leading to Eq.(\ref{eq4}),
experimental conditions are assumed ideal with 100$\%$ fringe
contrast of the green interference. Experimentally a contrast of
50$\%$ is observed, such that Eq.(\ref{eq4}) only accounts for
half of the measured correlations, while the remaining part
corresponds to $G^{(2)}_{ni}$. Therefore in all data sets for
$G_m^{(2)}(T)$ shown below, the measured $G^{(2)}_{ni}(T)$ has
already been subtracted from the raw histogram data.

\begin{figure}
\includegraphics[width=6.5cm]{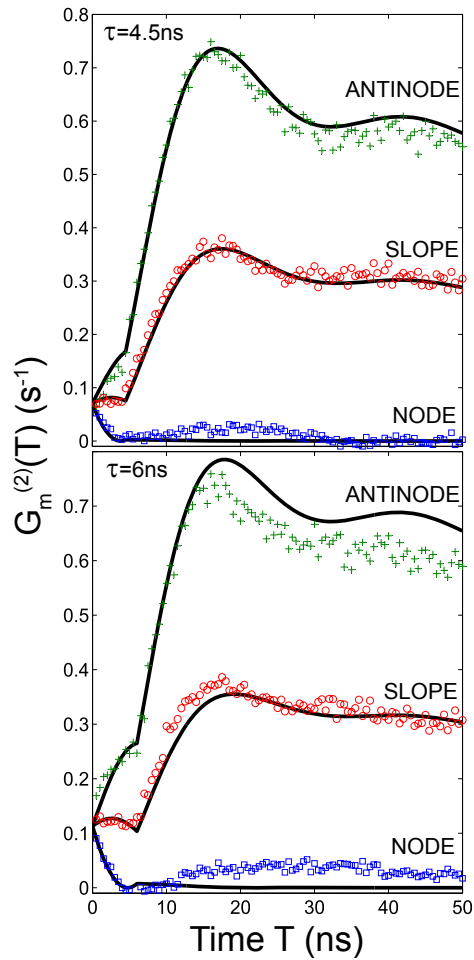}
\caption{Measured correlation function $G_m^{(2)}(T)$, after
subtraction of the non-interfering part, for the ion placed near a
node (squares), a slope (circles) and an anti-node (crosses) of
the standing-wave mirror mode. Each data set corresponds to 3
hours of integration. The lines represent the results of our model
(Eq.(3)).} \label{fig3}
\end{figure}

Figure 3 presents such measured second order correlation functions
$G_m^{(2)}(T)$ for interfering ion and mirror image. We compare
three relevant situations: the ion close to a node
($k_{fl}L=0.03\pi$), on the slope ($k_{fl}L=0.28\pi$) and close to
an antinode ($k_{fl}L=0.4\pi$) of the standing-wave mirror mode.
The first notable feature is that always $G_m^{(2)}(0)>0$. For our
single trapped ion, such coincidence can only appear when a
directly emitted and reflected photon are simultaneously detected.
This is possible in our experiment since the delay of a reflected
photon is comparable to the time required to re-excite the ion to
the P$_{1/2}$ state. The second important feature is that
all situations show the same coincidence rate $G_m^{(2)}(0)$,
although the relative phase (2$k_{fl}L$) between the coincident
direct and reflected photon fields is different in the three
situations. This demonstrates that at $T=0$ one has the full
which-way information about the two photons. Consequently no
interference can be observed.

We now discuss the long-time limit $T \gg \tau$: in Eq.(\ref{eq4})
the time argument of $b_{{\rm P}_{1/2}}$ reduces to T and \newline
$G_m^{(2)}(T)$ $\equiv$ $\sin^4(k_{fl}L) |b_{{\rm P}_{1/2}}(T)|^2
|b_{{\rm P}_{1/2}}^{(ss)}|^2$, $|b_{{\rm P}_{1/2}}^{(ss)}|^2$
being the steady state population of the P$_{1/2}$ state. The
second order correlation function thus factorizes into the product
of the first order correlations at time $t$ and ($t+T$). For the
anti-node position, the interference is constructive and
$G_m^{(2)}(T\gg\tau)$ is maximal. On the other hand, at the node
position the fully established destructive interference suppresses
the detection of photon pairs with long time intervals between
them, thus creating a strong effective bunching around T=0 despite
the fact that we are dealing with only a single atom.

Finally we study the correlations for short time delay between
photon detections,\ $0<T\leq\tau$. In this regime memory effects
are crucial, as one can see from Eq.(\ref{eq4}), where
excited-state amplitudes at different times are superimposed. The
difference between the three positions originates mainly from the
weight $\cos(2k_{fl}L)$ of the first term in Eq.(\ref{eq4}), which
corresponds to the processes where both photons are emitted in the
same direction. The two other terms, describing processes where
they take opposite directions, do not depend on the mirror phase.
As a result, a conspicuous kink in all the curves at
$T\approx\tau$ is observed. This kink marks the sudden onset of
full interference, when no more which-way information is present.

To summarize, for a single ion trapped and laser-excited in front
of a mirror, we have presented the second order time correlation
function of emitted photons. Depending on the position of the ion,
e.g. at a node or at an antinode of the reflected field standing
wave, very different behaviours are shown for large distances
between the ion and the mirror. In this non-Markovian regime, the
detection of photon pairs separated by a large time interval is
modulated by the interference experienced by each photon. On the
other hand, coincident two photon detections are insensitive to
the exact position of the ion, because interference can not be
established and which way information for each detected photon is
accessible. Consequently, when the ion is placed at a node of its
reflected fluorescence standing wave, a single photon detection is
prohibited by first order interference while a joint two photon
detection is allowed. This appears as a bunched profile in the
correlation function which reveals the transient regime of the
field establishment in our half cavity interferometer. We believe
that our analysis characterizes the transient regime of cavity
quantum electrodynamics.

This work has been partially supported by the Austrian Science
Fund (project SFB15), by the European Commission (QUEST network,
HPRNCT-2000-00121, QUBITS network, IST-1999-13021, SCALA
Integrated Project, Contract No. 015714), by a travel grant of the
{\"O}AD (No. 3/2005), the Spanish MEC (No. HU2004-0015) and by the
"Institut f\"ur Quanteninformation GmbH."

\end{document}